\documentclass[reprint, amsmath,amssymb, aps]{revtex4-2}

\usepackage{times}
\usepackage{graphicx}
\usepackage{subfig}
\usepackage{epstopdf}
\usepackage{psfrag}
\usepackage{ae}
\usepackage{amsmath,amssymb}
\usepackage{float}
\usepackage[usenames]{color} 
\usepackage[dvipsnames]{xcolor}
\usepackage[normalem]{ulem}
\newcommand*{\sGL}{\sigma_{\scriptscriptstyle\rm GL}}
\newcommand*{\sSL}{\sigma_{\scriptscriptstyle\rm SL}}
\newcommand*{\sSG}{\sigma_{\scriptscriptstyle\rm SG}}

\newcommand*{\Scap}{S_{\scriptscriptstyle\rm CAP}}

\begin{document}
	
	\preprint{AIP/123-QED}
	
	\title{Hierarchical structured surfaces enhance the contact angle of the hydrophobic (meta-stable) state}

	\author{Iara Patrícia da Silva Ramos} 
	%\email{}
	\affiliation{Instituto de F\'isica, Universidade Federal
		do Rio Grande do Sul, Caixa Postal 15051, CEP 91501-970, 
		Porto Alegre, Rio Grande do Sul, Brazil}
	
	\author{Cristina Gavazzoni} 
	\email{crisgava@gmail.com}
	\affiliation{Instituto de F\'isica, Universidade Federal
		do Rio Grande do Sul, Caixa Postal 15051, CEP 91501-970, 
		Porto Alegre, Rio Grande do Sul, Brazil}
	
	\author{Davi Lazzari} 
	%\email{}
	\affiliation{Instituto de F\'isica, Universidade Federal
		do Rio Grande do Sul, Caixa Postal 15051, CEP 91501-970, 
		Porto Alegre, Rio Grande do Sul, Brazil}
	
	\author{Carolina Brito}
	%\email{carolina.brito@ufrgs.br}
	\affiliation{Instituto de F\'isica, Universidade Federal
		do Rio Grande do Sul, Caixa Postal 15051, CEP 91501-970, 
		Porto Alegre, Rio Grande do Sul, Brazil}
	
	\date{\today}

	\begin{abstract}
		
		The relation between wetting properties and geometric parameters of fractal surfaces are widely discussed on the literature and, however, there are still divergences on this topic. Here we propose a simple theoretical model to describe the wetting properties of a droplet of water placed on a hierarchical structured surface and test the predictions of the model and the dependence of the droplet wetting state on the initial conditions using simulation of the 3-spin Potts model. We show that increasing the auto-similarity level of the hierarchy -- called $n$ -- does not affect considerably the stable wetting state of the droplet but  increases its contact angle. Simulations also explicit the existence of metastable states on this type of surfaces and shows that, when $n$ increases, the metastability becomes more pronounced. Finally we show that the fractal dimension of the surface is not a good predictor of the contact angle of the droplet.
		
	\end{abstract}
	
	\maketitle
	
	\section{Introduction}
	
	When a droplet is placed on a solid surface, it displays different final configurations that depend on several factors such as droplet volume, composition, surface chemistry and surface geometry \cite{Quere2008}. Hydrophobic surfaces are typically associated with the Cassie-Baxter \cite{Cassie1944} (\textbf{CB}) state, in which we observe air pockets trapped underneath the droplet, with high contact angles between droplet and surface while hydrophilic surfaces are associated with the Wenzel \cite{we36} (\textbf{WE}) state, characterized by the homogeneous wetting of the surface, and lower values of contact angle.
	
	Surfaces with contact angles $\theta_c > 150^\circ$ and small contact angle hysteresis are said to be superhydrophobic and have been intensely studied due to their many technological applications such as  self-cleaning surfaces \cite{barthlott97, Blossey2003} and water purification\cite{xue14,chan09, pa15, gava2021}. In nature there are several biological materials presenting these properties \cite{barthlott97,feng02, cheng2005lotus, liu10}. These material often have a dual-scale topography which leads to the idea that hierarchical structures are the key to achieve superhydrophobicity.
	
	Many experimental, theoretical and computational studies were made on this field in order to establish a relationship between the structure and the wetting properties. Onda \textit{et. al.} \cite{Onda1996} proposed a theoretical model of wetting in fractal surfaces in which the contact angle of the droplet has a power law dependency on the fractal dimension of the substrate. They also performed experiments on AKD fractal and flat surfaces and showed that the fractality enhances the wetting properties of the flat surface. A subsequent study by Shubuichi \textit{et. al} on aluminum surfaces treated with several hydrophobic surfaces coupling agent also reach the same conclusion \cite{shibuichi1998super}. Synytska  \textit{et. al}  \cite{synytska2009wetting} performed simulations and experiments of different polar and no-polar liquids on fractal surfaces made of polymer- or silene - coated "core-shell" particles and found that their results deviate from the predictions made by Onda and Shubuichi. They hypothesize that the origin of this deviation could be associated with the appearance of metastable states of the droplet.
	
	Using molecular dynamics simulations of liquid droplets in contact with self-affine fractal surfaces, Yang \textit{et. al}\cite{yang2006influence} showed that the contact angle of the droplets are strongly dependent on the surface roughness but nearly independent of the fractal dimension. Further theoretical and experimental work reach the same conclusion \cite{jain2017fractal}.
	
	On the other hand, Gao \textit{et. al}\cite{gao2016tunable} fabricated several fractal hierarchical materials and found a clear dependence of the contact angle on the fractal dimension. Further studies on PET surfaces treated with cold oxygen plasma \cite{piferi2021hydrophilicity} concludes that the fractal dimension could be the most important predictor to predict surface properties.
	
	The examples above illustrate that, despite all work developed about fractal surfaces, the dependence of the wetting properties on the  geometric parameters and the fractal dimension are still not clear. In this work, we study theoretically and numerically the wetting properties of a particular model of fractal surfaces, namely a hierarchically structured surface.  From the theoretical point of view, the surface can have infinite similarity levels $n$. We assume a 3D spherical droplet which can be in two different wetting states when placed on this surface: one that homogeneously wets the surface and is referred as Wenzel state, and another that retains air below the droplet, called as a Fakir Cassie-Baxter state. Due to the hierarchy of the surface, we are able to compute the energy of both wetting states in any level $n$. Applying a minimization procedure, we predict the most probable wetting state and the contact angle of the droplet in any $n$. To test the models predictions and the dependence of these states on the initial wetting condition of the droplet, we employ Monte Carlo simulations of the three states Potts model. Finally we discuss our results in terms of the fractal dimension obtained for our surfaces.
	
	This manuscript goes as follow. First, in section \ref{theoreticalModel} we introduce our theoretical model and discuss the predictions of the model. In section \ref{section_simu} we present details on the 3 spin Potts model used in this work and in section \ref{section_result} we show and discuss our results. Finally in section \ref{section_conclu} we summarize our conclusions.

	\section{Theoretical Continuous Model}
	\label{theoreticalModel}
	
	In this section, we develop a theoretical model to determine the wetting state and the contact angle of a water droplet placed on top of the fractal surface. The model takes into account the global energy to create interfaces as used in several works \cite{Quere2008, Sbragaglia2007, Tsai2010, Shahraz2012, Lopes2013} and then we apply a minimization process to define the lowest energy wetting state first proposed by Fernandes \textit{et. al.} \cite{fernandes2015} and also applied in other studies \cite{Silvestrini2017, laz19, gava2021}.
	
	For this purpose, we assume that a droplet with fixed volume $V_0=4/3\pi R_0^3$ can be found in two distinct wetting states, one called Wenzel (W) and characterized by the homogeneous wetting of the surface and one called Cassie-Baxter (CB) with air pockets trapped underneath the droplet. The Wenzel and Cassie-Baxter states are associated with hydrophilic and hydrophobic behavior respectively.
	
	\begin{figure*}
		\includegraphics[width=\textwidth,height=3cm]{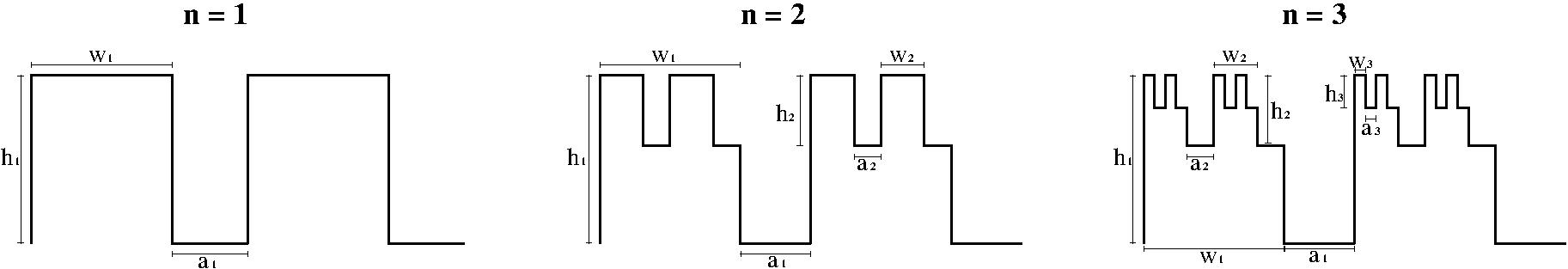}
		\caption{Lateral view of the pillared fractal surface for different levels of auto-similarity $n$ and definition of the geometric parameters. This representation is for $m=2$.}
		\label{geometria}
	\end{figure*}
	
	The energetic cost associated with each of these states is given by $E^s_{\rm tot} = \Delta E^s + E_g^s$, where superscript $s$ represents the wetting state (CB or W) and $E_g$ is the gravitational energy. $ \Delta E^s$ is the difference in the interfacial energy between every pair formed from different phases (liquid, solid, and gas) after the droplet is placed on the surface in state $s$ and the energy of the surface without the droplet. The importance of gravitational energy depends on the droplet's size and its composition. In this work, we consider droplets with small volumes such that gravitational energy is negligible compared to $\Delta E^s$. % \cite{fernandes2015}. 
	
	The interfacial term  $\Delta E^s$ depends on the geometry of the substrate. In this work we explore a spherical droplet placed on a pillared fractal surface in several levels of auto-similarity, as exemplified in Figure \ref{geometria}. Each level is defined by a pillar height $h_n$, an interpillar distance $a_n$, and pillar width $w_n$, where the subscript $n$ stands for the auto-similarity level. For $n>1$ we also define a quantity %$m_n^2$ 
	$m$ that represents the linear number of pillar in the $n$-{th} level
	
	\begin{equation}
		m_n = \frac{w_{n-1}}{d_n} 
		\label{m}
	\end{equation}
	
	\noindent where $d_n = a_n + w_n$, and the total number of new smaller pillars generated above the "$n-1$"-pillar level is $m_n^2$. Figure \ref{geometria} shows an example for $m=2$ and define the geometric parameters. The droplet is supposed to have a spherical cap with volume $V_0$, radius $R$, height $H$, base radius $B$ and it touches the fractal pillared surface with a contact angle $\theta_c$. 
	
	Our equations and results are represented in terms of the aspect ratio $A_n = {h_n}/{w_n}$ and the local solid fraction  $\phi_n={w_n^2}/{d_n^2}$. $A_n$ can assume any positive real value and $\phi_n$ assume values between $0$ and $1$.
	
	The interfacial energy equations for pillared fractal substrate in the $n-th$ auto-similarity level  is given by
	
	\begin{eqnarray}
		\Delta E_{n}^{CB} &=&\sigma_{LG} \left[ \Scap^{CB} + \pi B_{CB}^{2}\left( 1 - \prod_{i = 1}^{n} \phi_i( 1 + \cos{\theta_Y}) \right) \right], \ \ \label{Ecb}\\
		\Delta E_{n}^{W} &=& \sigma_{LG}\left [ \Scap^{W} - \pi B_{W}^2\left( 1 +  S_{n}\right)\cos{\theta_Y}\right], \label{Ewe}
		%S_n &=& 4A\left[ \sum_{i=1}^{n}\left(\prod_{j = 1}^{i} \phi_{i}\right) -  \sum_{i=1}^{n-1}\left(\prod_{j = 1}^{i} \phi_{i}\right)\frac{h_{i+1}}{h_i}\right] \label{Sn}
	\end{eqnarray}
	
	\noindent where $S_n = 4\left[ \sum_{i=1}^{n}\left(\prod_{j = 1}^{i} \phi_{j}\right)A_i -  \sum_{i=1}^{n-1}\left(\prod_{j = 1}^{i} \phi_{j}\right)A_i\frac{h_{i+1}}{h_i}\right]$. For all equations above $\Scap^s=2\pi {R_s}^2[1-\cos(\theta^s_c)]$ is the surface area of the spherical cap, $B_s=R_s \sin(\theta^s_c)$ is the droplet base radius and $\theta^s_c$ is the contact angle of the droplet in the state $s$. The solid-gas surface tension, $\sGL$, and the Young's contact angle, $\theta_Y$ are chosen to be in agreement with experiments with water on poly-(dimethylsiloxane) (PDMS) surface \cite{Tsai2010}: $\sGL= 70$mN/m and $\theta_Y = 114^{\circ}$.
	
	%We note that equations \ref{Ecb} and \ref{Ewe} are general for any self-similar square pillared surface with any combination of $\phi_n$ and $A_n$. It can be shown that these equations are general for any polygon shaped pillared surface  by  substituting the constant $4$ in the term $S_n$ %Eq. \ref{Sn}  by a geometric function, \cbb{as shown for some examples in the SM}. \cris{TENHO QUE CONFIRMAR ISSO AQUI}. \cbcom{CHECAR E DEIXAR exs no SM.}
	
	In this work we focus on the special case were $A_1=A_2= \ldots =A$, $\phi_1=\phi_2=\ldots=\phi$, $m_2=m_3=\ldots=m$, and the surface is so-called a self-similar surface. In this particular case equations \ref{Ecb} and \ref{Ewe} can be simplified as
	
	\begin{eqnarray}
		\Delta E_{n}^{CB} &=&\sigma_{LG} \left[ \Scap^{CB} + \pi B_{CB}^{2}\left( 1 - \phi^n( 1 + \cos{\theta_Y}) \right) \right] \label{Ecbn}\\
		\Delta E_{n}^{W} &=& \sigma_{LG}\left [ \Scap^{W} - \pi B_{W}^2\left( 1 + S_{n}\right)\cos{\theta_Y}\right] \label{Ewen}%\\
		%S_n &=& 4A \left[ (\sum_{i=1}^n \phi^i) - \frac{\sqrt{\phi}}{m}\sum_{i=2}^n \phi^{i-1} \right] 
		\label{eqsAphi}
	\end{eqnarray}
	
	\noindent where $S_n = 4A \left[ (\sum_{i=1}^n \phi^i) - \frac{\sqrt{\phi}}{m}\sum_{i=2}^n \phi^{i-1} \right]$. 
	
	%It is \sout{easy to see} possible to show that that for $n \rightarrow \infty$ the energy associated with the wetting state CB will only depend on the geometry of the droplet while the energy for the W state will depend on the geometry of the surface as well.  This is shown in further detail in the supplementary material. \cbcom{Precisamos desta frase? Se sim, eh importante guiar o leitor para entender o que queremos dizer usando os termos das eqs. Assim como esta nao esta clara e nao me parece ser necessaria pro que vamos discutir.}
	
	%\cbcom{Lembro que precisavamos escolher o $h_i$ como parametro livre, mas pelas eqs acima isto nao aparece mais.. onde ficou esta dependência ?} \cris{Precisamos definir o h1 por causa da expressão do volume de baixo da gota.} \cbcom{as equacoes do volume devem estar no SM e ali o h1 deve ser mencionado.}
	
	\subsection{Model predictions}
	
	We now discuss two predictions of the model described by Eqs. \ref{Ecbn} and \ref{eqsAphi}. The first one is the identification of  the favorable wetting state from the thermodynamic point of view for these surfaces.
	To do so we find the global minimum energy state of the Eqs.  \ref{Ecbn} and \ref{Ewen}. This energy minimization process was introduced in \cite{fernandes2015} and goes as follow: first, we fix the surface parameters ($A$, $\phi$, $m$, $n$ and $h_1$(see SM)) and the volume  $V_0=\frac{4}{3}\pi R_0^3$  of the droplet. Next, we solve a cubic equation to obtain the radius of the droplet, $R^s$, for each state, CB or W; and lastly, we vary the contact angle $\theta_c^s \in [0, 180^{\circ})$ and use Eq.(\ref{Ecbn}) to obtain  $\Delta E^{\rm CB}$ and Eq.(\ref{Ewen}) for $\Delta E^{\rm W}$. This process finds the global minimum energy for CB and W states (respectively, $\Delta E^{\rm CB}_{\rm min}$ and $\Delta E^{\rm W}_{\rm min}$) and defines the thermodynamic wetting state, the one with the lowest $\Delta E_{\rm min}$.

\begin{figure}
		\centering
		\includegraphics[width=0.7\linewidth]{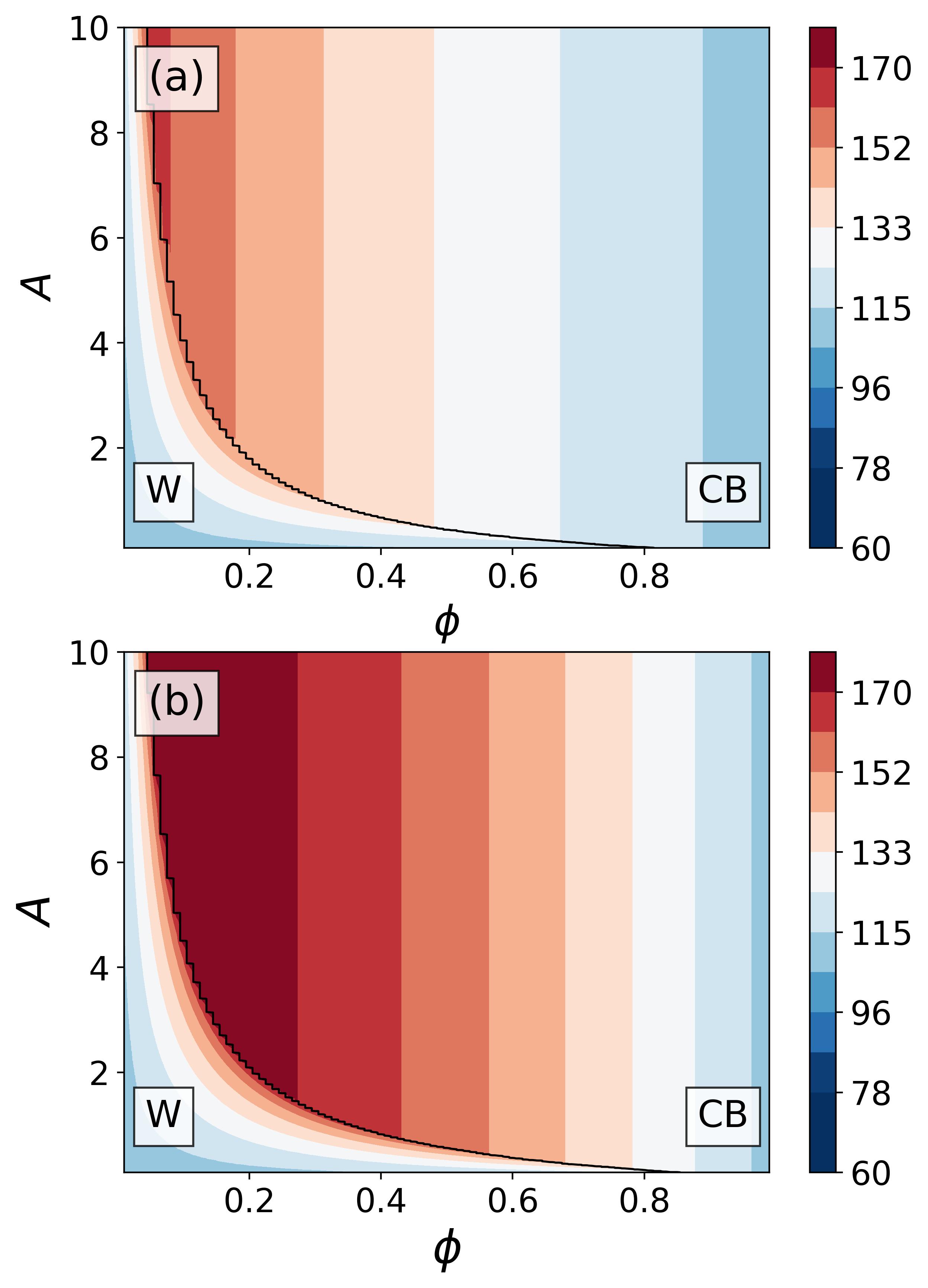}
		\caption{Wetting diagram for $R \rightarrow \infty$, $m=3$, (a) $n=1$ and (b) $n=3$. Colors represent the contact angle and the solid line marks the transition from the W state to the CB state.}
		\label{diagrama}
	\end{figure}

	This minimization process is employed in order to build a theoretical wetting diagram for the fractal surface in different auto-similarity levels as a function of $A$, $\phi$, $R_0$ and $m$. Figure \ref{diagrama} shows this diagrams for fixed parameters $m=3$, $R_0\rightarrow \infty$, $n=1$ and $3$, and we screen over $A\in (0,10]$ and $\phi \in (0,1]$. 
	
	\begin{figure*}
		\centering
		\includegraphics[width=\textwidth,]{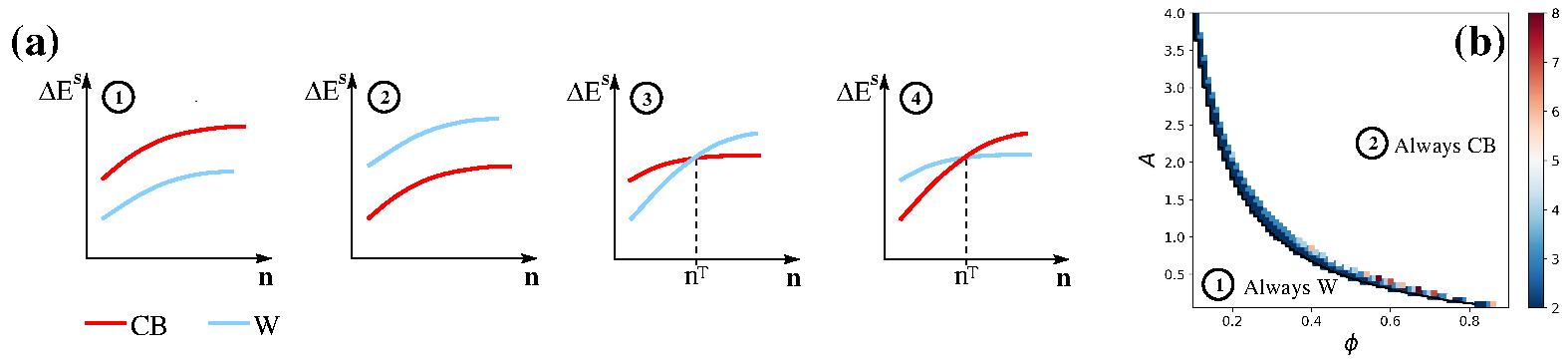}
		\caption{(a) Schematic of all outcomes predict by the theory for $R\rightarrow \infty$ (b) Transitional diagram. Color represents the auto-similarity level $n^T$ where a transition occurs from the original wetting state for $n=1$. }
		\label{trans}
	\end{figure*}
	
	Fig. (\ref{diagrama}) shows the theoretical wetting diagram as a function of $A$ and $\phi$. Fig. \ref{diagrama}(a) is the diagram at the level $n=1$ and Fig. \ref{diagrama}(b) at the level $n=3$. These figures indicate a non trivial dependence of the most  favorable wetting state  on the geometric parameters of the surface: small $\phi$ and $A$ favors the W state. When $A$ increases, CB becomes the most favorable wetting state, which is in agreement with other works \cite{Shahraz2013, Lundgren2003}. 
	Color-code represents the value of the contact angle $\theta_c$. One observes that $\theta_c$ presents higher values in the CB region compared to W region, in particular close to the transition line --  shown in a solid black line. Comparing the diagram for the case $n=1$, Fig. \ref{diagrama}(a), with $n=3$, Fig. \ref{diagrama}(b), one observes that the transition line does not change drastically when increasing the auto-similarity level $n$. In other words, if the favorable wetting state is W, for $n=1$, increasing the auto-similarity level does not change the favorable state. On the other hand,   increasing $n$ leads to CB states with higher values of contact angle, \textit{i.e.}, the surface becomes more hydrophobic. We studied the dependence on the droplet size $R_0$ (shown in the SM) and identified that these observations remains the same for all values of $R_0$.

	We now evaluate how the energy of the wetting states CB and W changes when the auto-similarity level $n$ increases and in which cases the increase in the energy level  favors the CB or the W state. Considering droplets with $R\rightarrow \infty$ (physically it means that the radius of the droplet is much bigger than the typical sizes of the substrate textures) one can rewrite the energy of the droplet in a given auto-similarity level $n+1$ as a function of the previous level $n$. Let us first evaluate the CB case, Eq. \ref{Ecbn}:
	\begin{eqnarray}
		\Delta E_{n+1}^{CB}   &=&  \Delta E_{n}^{CB} +\sigma_{LG}\pi B^{2} \left( 1 -\phi\right) \left(1 - M^{CB}\right), \label{Ecbn+1}\\
		M^{CB} &=& 1 - \phi^n( 1 + \cos{\theta_Y}), \nonumber
	\end{eqnarray}
	note that $\Delta E_{n+1}^{CB}  >  \Delta E_{n}^{CB}$  when  $\left( 1 -\phi \right) \left(1 - M^{CB}\right) > 0$ and smaller otherwise.

	Since $ 0 \leq \phi \leq 1$, the term $(1 -\phi)$ in Eq. \ref{Ecbn+1} 
	is positive or zero in the case $\phi=1$.  For the latter case, no dependence of $\Delta E^{CB}$  is expected when changing the level $n$. Also, for hydrophobic surfaces considered in this work, one has $-1 <\cos(\theta_Y) <0$, which implies that $0 < M^{CB} < 1$ and leads to $(1 - M^{CB})>0$. Furthermore, it is interesting to note that as $n$ increases, $\phi^n \rightarrow 0$ and then $M^{CB} \rightarrow 1$. Therefore, the energy associated to the wetting state CB increases with the auto-similarity level $n$, $E_{n+1}^{CB}  > \Delta E_{n}^{CB}$,  until it reaches a plateau. 
	
	In the case of the W state, the interfacial energy in the $(n + 1)$ level can be written as:
	\begin{eqnarray}
		\Delta E_{n+1}^{W} &=& \Delta E_{n}^{W} - 4\sigma_{LG} \pi B^2  M^{W}cos{\theta_Y} ,\label{Ewen+1}\\
		M^{W} &=& A\phi^{n} \left[ 1 - \phi \frac{\sqrt{\phi}}{m}\right], \nonumber
	\end{eqnarray}
	thus, for hidrophobic surfaces, where $\theta_Y>90^{\circ}$, since $ 0 \leq \phi \leq 1$, $A > 0$ and $m \geq 1$$, \Delta E_{n + 1}^{W} > \Delta E_{n}^{W}$ for any $\phi$ and $m$. Also, similar to the CB case, as $n$ increases $M^{W} \rightarrow 0$, for $\phi \neq 1$, and the energy associated with the W state goes to a plateau.

	These results lead to an interesting conclusion: if a droplet is found, for instance, on the W state on the \textit{n-th} level, a transition for the CB state on the level $n+1$ will occur only if the energy associated with the W state grows faster with $n$ than the energy for CB. The rates at which each one of these energies grows depends on the geometric parameters of the surface. A schematic figure of all the possible outcomes predicted are shown in Figure \ref{trans}(a). Case 1 and 2 presents no transition between the states when level $n$ changes, while cases 3 and 4 represents geometries where the droplet changes its wetting state as a function of $n$: for example in the case 3 the thermodynamic favorable state is W for $n<n^T$ and CB when $n>n^T$. 
	
	Figure \ref{trans}(b) is referred as {\it transitional diagram} and it shows all the combinations of $A$ and $\phi$ where there is a transition from one wetting state  to the other at a given $n^T$. The color represents the value of $n^T$ where this transition occurs. This transition as the auto-similarly level increases only occurs in a small region  of the $A\times\phi$ diagram and it is restricted to the vicinity of the wetting phase transition for $n=1$. This leads to the conclusion that increasing the auto-similarity level does not propitiate a change in the wetting behavior of the surface (cases 1 and 2 shown in Fig. \ref{trans}(a)) except near the transition line where $E_n^{CB} \approx E_n^{W} $ (cases 3 and 4 in Fig. \ref{trans}(a)).  
	
	All together, our theoretical model indicates an increase of $\theta_c$ when the similarity level $n$ increases, as shown in Fig. (\ref{diagrama}). This observation is in line with the 
	enhancement of the hydrophobic behavior in fractal structures reported in several studies \cite{gao2016tunable,piferi2021hydrophilicity,shibuichi1998super}. Our results also support that a change in the wetting behavior of the surface when the auto-similarity level increases only happens in a narrow region close to the transition line, as shown in the Fig. \ref{trans}(b).
	
	One limitation of this method, however, is that it is not able to capture metastable states. Metastability in wetting is a well know phenomena\cite{Quere2008} and has been reported in several studies \cite{fernandes2015,laz19,AMI_Marion2021}. This may be the reason why some authors report a change in the wetting behavior when increasing fractality \cite{hazlett1990fractal,Kao1996,shibuichi1998super}.  In some experiments the metastability is manifested through the dependence of the final wetting state on the droplet's initial condition \cite{Quere2008, Koishi2009}.  To test this condition we implement Monte Carlo (MC) cellular Potts model as explained in the next section. 
	
	\section{Simulations: 3-spins Cellular Potts model}
	\label{section_simu}
	
	Monte Carlo simulations (MC) of the cellular Potts model are widely used to study wetting phenomena in textured surface \cite{Lopes2013,Oliveira2011,Mortazavi2013, fernandes2015, gava2021}. This method allows to simulate bigger droplet sizes compared to other numerical approaches used to explore hydrophobic surfaces, such as molecular dynamics (MD)\cite{Koishi2011,Wu2009} and lattice Boltzmann methods \cite{Dupuis2005, Sbragaglia2007} and thus, it is ideal for treating mesoscopic systems and, therefore, more appropriate for comparison with experimental results.
	
	The idea of the model is that both the tree dimensional droplet and the surface are divided in cubes with lateral size $p$. To each cube is associate a state $s_i$ that represents one of the components of the system: gas, water or solid. The Hamiltonian is then given by:
	\begin{eqnarray}
		H &=& \frac{1}{2} \sum_{\langle{\rm i},{\rm j}\rangle} E_{s_{\rm i},s_{\rm j}}(1-\delta_{s_{\rm i},s_{\rm j}}) + \alpha \left( \sum_{\rm i} \delta_{s_{\rm i},1}-V_T \right) ^2 + \nonumber \\
		& & g \sum_{\rm i} m_{\rm i} h_{\rm i} \delta_{s_i,1}  ,
		\label{hamil}
	\end{eqnarray}
	where the spin $s_i \in \{0,1,2\}$ represent gas, water and solid states, respectively.
	
	The first term in Eq. (\ref{hamil}) represents the energy related to the presence of interfaces between sites of different types. The summation ranges over pairs of neighbors, which comprise the 3D Moore neighborhood in the simple cubic lattice (26 sites, excluding the central one), $E_{s_i,s_j}$, are the interaction energies of sites $s_i$ and $s_j$ of different states at interfaces, and  $\delta_{s_i,s_j}$ is the Kronecker delta. 
	
	In the second term in Eq. (\ref{hamil}), $V_T$ is the target volume, the summation is the instantaneous droplet volume and the parameter $\alpha$ mimics the compressibility of the liquid. Thus, this term maintains the droplet volume constant during the simulation. The last term is the gravitational energy, where $g = 10m/s^2$ is the acceleration of gravity and $m_i$ is the mass of the site. In both the volumetric and gravitational terms, only sites with $s_i = 1$ contribute.
	
	The parameters for the Hamiltonian in Eq. \ref{hamil} were based on those used in experiments with water on poly-(dimethylsiloxane) (PDMS) surface \cite{Tsai2010}: $\sGL= 70$mN/m, $\sSG = 25 $mN/m and the solid-liquid surface tensions were obtained from Young’s relation $\sGL \cos(\theta_Y) = \sSG - \sSL$, where ${\theta_Y}$  is the contact angle on a smooth surface and assumes the values $\theta = 114^{\circ}$. These values are divided by 26, which is the number of neighbors that contributes to the first summation of our Hamiltonian. The length scale of our simulation is such that one lattice spacing corresponds to $p\mu$m, where $p=0.20$, therefore, the interfacial interaction energies are given by $E_{s_i,s_j} = A\sigma_{s_is_j}/26$, with $A=p^2\mu$m$^2$. This choice results in the following values for the interfacial energies:  $E_{0,1} = p^2 2.70 \times 10^{-9}\mu$J, ~$E_{0,2} = p^2 0.96 \times 10^{-9}\mu$J, ~$E_{1,2} = p^2 1.93\times 10^{-9}\mu$J. The mass of water existent in a unit cube is $m_w= 10^{-15}$kg and $\alpha = 0.01 \times 10^{-9}\mu$J/($\mu$m$)^6$.
	
	The total run of a simulation is $1 \times 10^{6}$ Monte Carlo steps (MCS), from which the last third are used to measure observables of interest. Each MCS is composed of $V$ number of trial spin flips, where $V$ is the number of water sites. A spin flip is accepted with probability $ \text{min}\{1,\text{exp}(- \beta \Delta H)\} $, where $\beta = 1/T$. In the cellular Potts model, $T$ acts as noise to allow the phase space to be explored. In our simulations, a value of $T=13$ was used, which allows an acceptance rate of approximately 22$\%$. 
	
\begin{figure}
		\centering
		\includegraphics[scale=0.35]{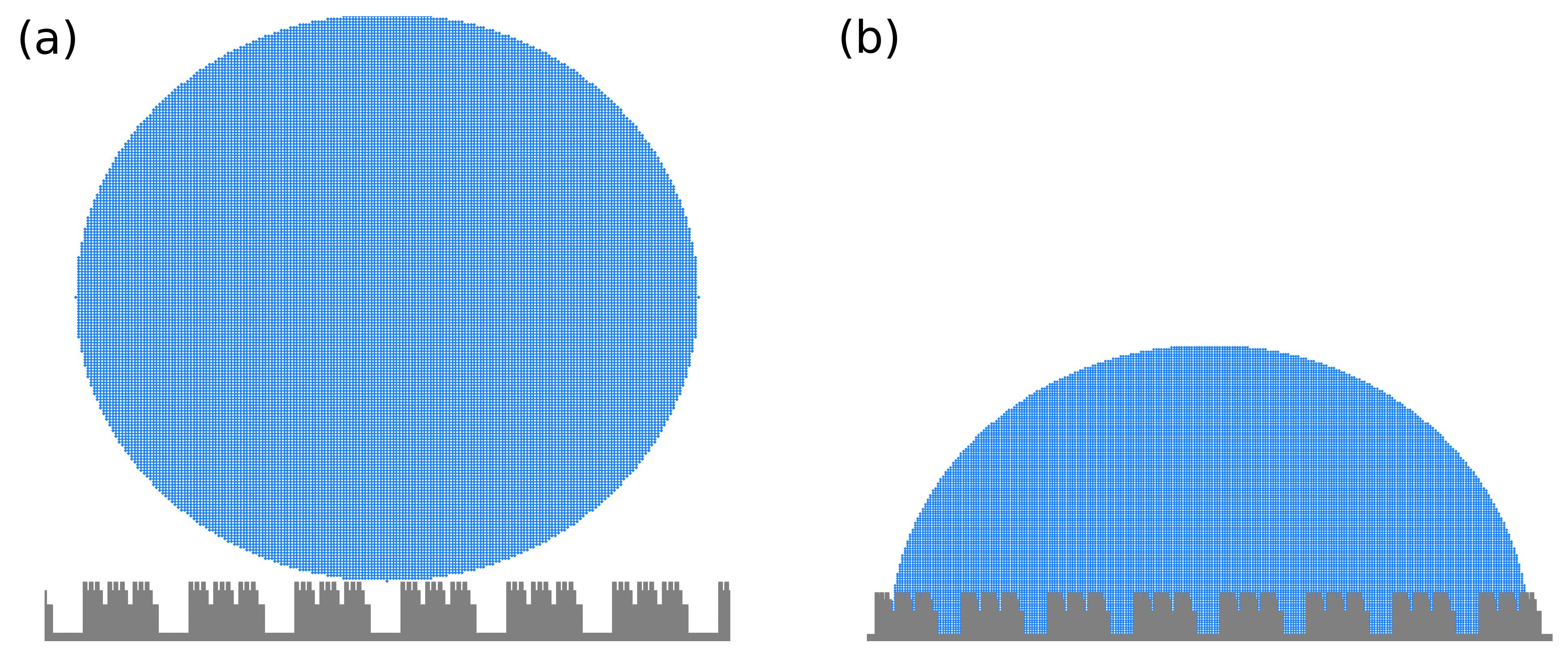}
		\caption{Schematic representation of the simulations initial condition.: (a) CB$^0$ and (b) W$^0$}
		\label{CI}
	\end{figure}

	As mentioned previously, a relevant aspect of the wetting problem is the metastability of the wetting states. In these cases, even though a given state is the lowest energy, the droplet does not reach it and displays another state.  The simulations allow us to study if the final wetting state of the droplet depends on the initial state that the droplet is placed on the substrate. To test this dependence, the droplet is initialized in two different wetting regimes. One possible wetting initial state is exemplified in Fig. \ref{CI}(a): it consists of a spherical droplet with 
	$V  \approx V_0=4/3 \pi R_0^3$ placed on top of the pillars. Because the droplet does not fill the surface, we refer to this as an initial Cassie-Baxter state and call it CB$^0$. The second initial state is shown in  Fig. \ref{CI}({b}) and corresponds to a  hemisphere which wets the substrate homogeneously and has the same initial volume $V_0$. We refer to it as an initial Wenzel state and identify it with  $W^0$.
	
	Through this work, it is used $R_0=20\mu$m in a cubic box with $L=(400p)\mu m$. The substrate was construct in a way that $A_1=A_2= \ldots =A$, $\phi_1=\phi_2=\ldots=\phi$ and $m_2=m_3=\ldots=m=3$.  Due to the discreteness of the lattice the values for $A$ and $\phi$ for the final structure are approximated. For more details see the SM.

	\section{Results and Discussion}
	\label{section_result}
	
	In this section, we discuss our simulations results for  different auto-similarity levels $n$, $m = 3$, $A\in [0.5,4.0]$ and $\phi \in [0.1,0.6]$. Due to numeric resources limitations and the need to span a big range of parameters and different hierarchical levels, we use a droplet of initial radius $R_0 = 20\mu m$. Also, the highest level that we are able to simulate is $n=3$. First we compare our simulations with the theoretical predictions discussed in previous section and then check the metastability of the wetting states as a function of $n$.  
	
	\begin{figure*}
		\centering
		\includegraphics[width=0.9\textwidth]{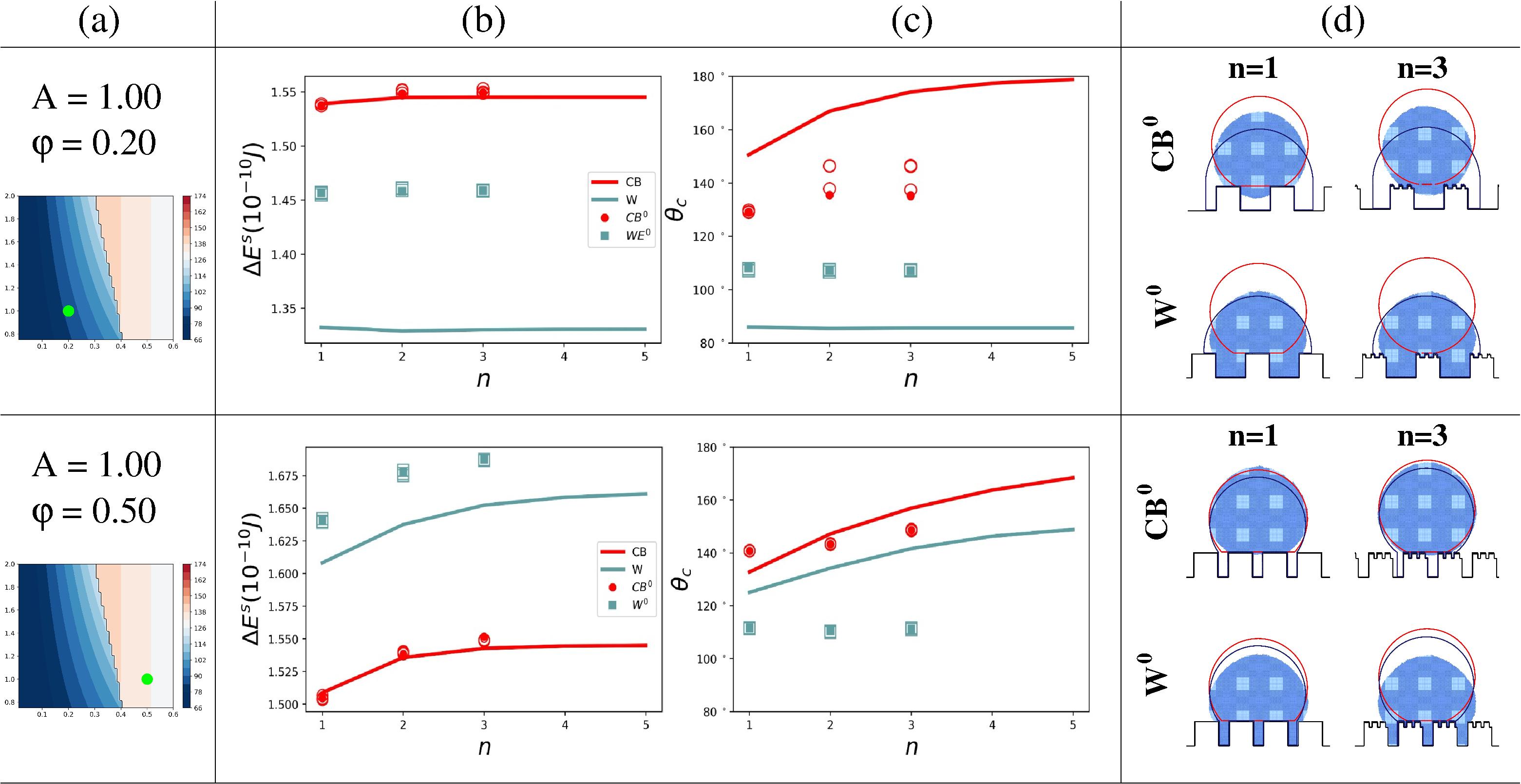}    
		\caption{{\bf Theory {\it vs} simulations} (a) Piece of the wetting diagram to indicate in green the geometrical points analyzed in this figure.
			The data shown in the line above correspond to $A=1, \phi=0.2$ which is a point in the W phase. Data shown in the line below is for $A=1, \phi=0.5$, corresponding to a point predicted to be in the CB phase.  
			(b) Energies $\Delta E^s$ and (c) contact angles $\theta_C$ as a function of the auto-similarity levels $n$.  Solid lines represent the theoretical results obtained from the minimization of equations \ref{Ecbn} (red) and \ref{Ewen}(blue), the circles represent the results of our simulation initializing the system from the CB$^0$(red) and squares the droplet initialized in W$^0$(blue). 
			(d) Cross sections of the droplet configurations in the final state of the Monte Carlo simulation for the points shown in solid symbols in (b) and (c). Above the simulation starts from the CB$^0$ configuration and below from the W$^0$ configuration. The solid (blue) line represents the cross section for the minimum-energy W configuration, and the dashed (red) line, the one for the minimum-energy CB configuration. The curves and the snapshots correspond to droplets with $R_0 = 20 \mu m$ placed on a surface with varying $n$ and $m=3$. }
		\label{comp_teo}
	\end{figure*}
	
	According to equations \ref{Ecbn+1} and \ref{Ewen+1} for infinite droplet sizes  both the CB and W energies are expected to increase with $n$,  as discussed in the section \ref{theoreticalModel}.  Also, from the wetting diagram \ref{diagrama} we observed that in the CB region the contact angle of the droplet increases with the increase of the $n$. A similar diagram for $R_0=20\mu$m is shown in the supplementary material and the same qualitative behavior is observed.
	
	Figure \ref{comp_teo} shows a comparison between the theoretical predictions for $R_0=20\mu$m (solid lines) and the simulation results (circles and squares). We consider two points of the diagram to discuss in more detail: one that is predicted to be in the W phase ($A=1, \phi=0.2$) and another in the CB phase ($A=1, \phi=0.5$). Figure \ref{comp_teo}(b) shows the energy of both states and  their contact angle in Figure \ref{comp_teo}(c)  as a function of $n$. Note that for the point $A=1, \phi=0.2$, the theoretical  energy of the CB state is bigger than the energy of the W state, indicating that the W is the stable state for all values of $n$. It corresponds to the case 1 -- "always W" --  sketched in Figure (\ref{trans}).  For the case $A=1, \phi=0.5$ the opposite behavior is observed: the energy of the CB state is smaller for all $n$, corresponding to the situation 2 -- "always CB" -- sketched  in Figure (\ref{trans}).
	
	In regards to the energy, we observe a good qualitative agreement with theoretical predictions, showing an increase of the energy with the auto-similarity level $n$. An exception is for the W case in the point $A=1, \phi=0.2$, where the theory predicts a slight decrease with $n$ (possible due to the finite size of the droplet).  An exact agreement of the energy between theory and simulations is not expected to happen because of the discretization of the droplet assumed in the simulations.
	
	In both cases and for both initial conditions, the theoretical predictions for the contact angle present a more pronounced variation with $n$ than measured in simulations. However, for $A=1.00$ and $\phi=0.50$ e see that the contact angles obtained with the simulations starting in CB$^0$ show good agreement with the theoretical predictions. Moreover, it is observed an important dependency of the final wetting states on the initial state of the simulation: if initiated in the W$^0$, the final contact angles are smaller than if the initial state is CB$^0$. The disagreement between simulations and the value predicted by the theory shows that the droplet becomes trapped in meta-stable states instead of reaching the global minimum. 
	
	Figure \ref{comp_teo}(d) shows the  droplet configuration in the final state of the Monte Carlo simulation for the points shown in (b) and (c) for $n=1$ and $n=3$. This figure also shows the theoretical predicted droplet in CB (solid red line) and W (solid blue line). On both cases we observe the dependence on the initial conditions of the simulations. Note, however, that for $\phi=0.20$ the final configuration for the W$^0$ deviates from the W prediction from the theory. On the other hand, for $\phi=0.50$, where the stable state predicted by the theory is CB it is observed a good agreement with the final configuration obtained starting in CB$^0$. This is interesting because, despite the final simulated configuration for $\phi=0.20$ be found on a W state it is not the same minimum W state predicted by the theory, i.e., it is also a metastable state. These metastable state of the W phase are reported in other studies \cite{AMI_Marion2021}.
	
\begin{figure}[H]
		\centering
		\includegraphics[width=0.9\linewidth]{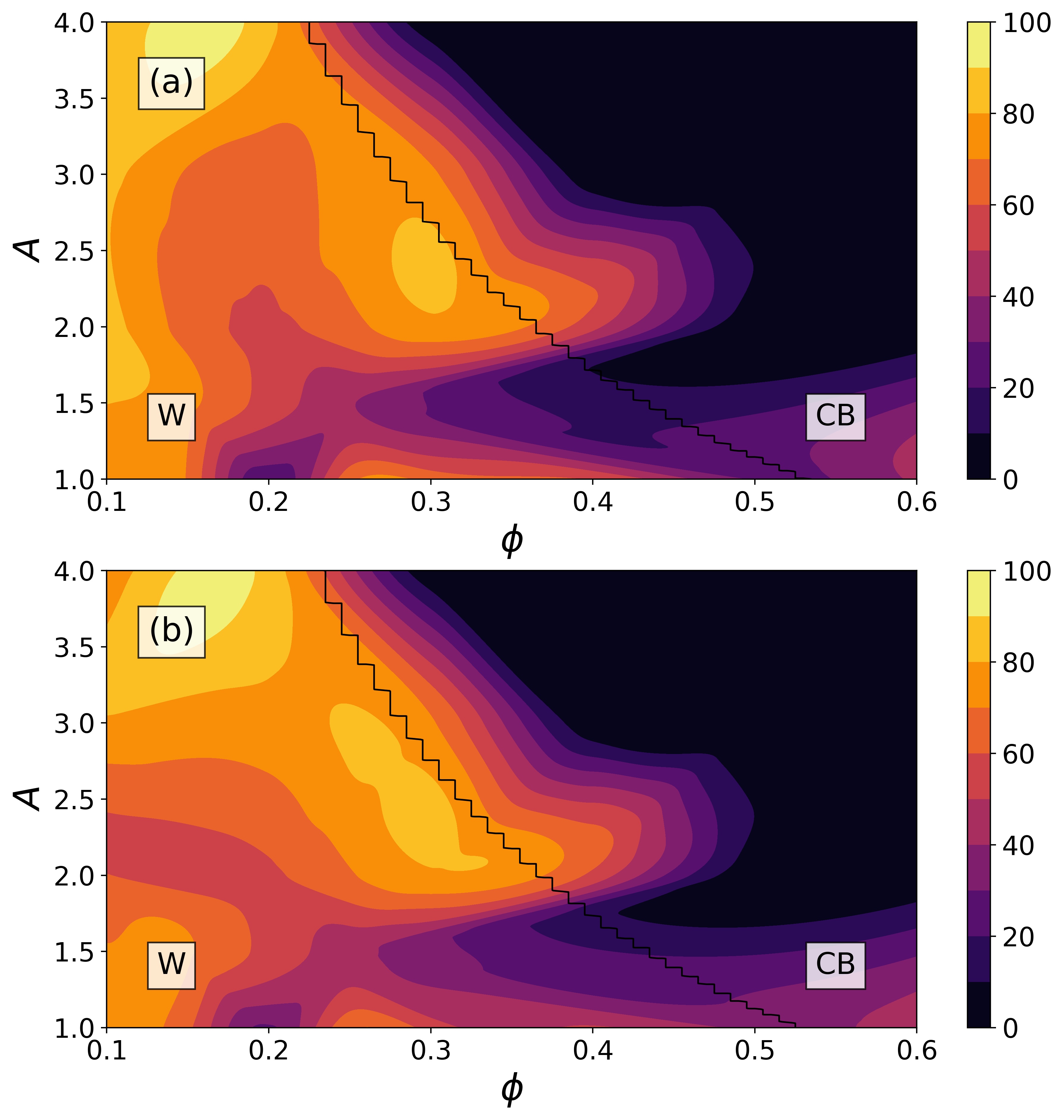}
		\caption{ $\Delta \theta$ as a function of $A$ and $\phi$ for (a) $n=1$ and (b) $n=3$. The color represent the values of $\Delta \theta$ and the solid line represents the theoretical transition line between the W and CB state for each auto-similarity level.}
		\label{colormap}
	\end{figure}

	In order to evaluate more systematically the metastability for all simulated points we measure the difference between the contact angles obtained with the different initial condition, $\Delta \theta = |\theta_{CB^0} - \theta_{W^0}$|. This is shown in Figure \ref{colormap} as a function of the geometric parameters $A$ and $\phi$ for $n=1$ and $n=3$. If $\Delta \theta = 0$ both simulations arrived to the same final state and so, no metastability is observed.
	
	For high values of $A$ and $\phi$, the W state becomes so costly energetically that even when the initial configuration is W$^0$, the final configuration is at the CB state. Then it is  observed that both initial conditions reach the same state and no metastability is observed.  When approaching the transition line from the CB phase, metastability is observed with $\Delta \theta$ varying from $50^{\circ}$ to $70^{\circ}$.
	
	The W region, on the other hand,  presents metastability for all the geometric parameters $A$ and $\phi$ for both $n=1$ and $n=3$. In other words, the droplets starting at different initial conditions gets trapped in different local minima, resulting in high values of $\Delta \theta$. Interestingly, there is a tendency of increase of $\Delta \theta$ as we increase the auto-similarity level $n$. Therefore, one can measure a very high contact angle even on the region were the W state is favorable and, increasing the auto-similarity level $n$ makes this metastable CB state more hydrophobic. This can be a possible explanation on why some authors see a change in the wetting properties when increasing fractality \cite{}. 
	
	The CB metastable state is a well known phenomena in wetting and was observed experimentally \cite{McHale2005, Quere2008, Liu2011}  and in computer simulations\cite{Shahraz2013, fernandes2015, AMI_Marion2021}. This metastability is related to a high barrier between states proportional to the height $h$ of the pillars (related to $A$) and it is highly dependent on how the droplet is placed on the surface. In regards to WE metastable state, a recent study by Silvestrini \textit{et. al.}\cite{AMI_Marion2021} calculated the free energy of a water droplet placed on a pillared surface found that several W metastables states are possible, with distinct contact angles. This is in agreement with our findings shown in fig. \ref{comp_teo}(c)-(d).
	
	We note that our results are in line with the experimental works as we now stress.  In experiments it is not possible to access the energy of the droplet; however the contact angle is measured and it is observed that its is enhanced in the case of  hierarchical structures \cite{Onda1996, Bhushan2009, Boreyko2011,kwon2018molecular} and more generally in fractal surfaces \cite{Onda1996}. Then it is not possible to know in most cases if the hydrophobic state is metastable or not. 
	
	In this work what we observe is  that the contact angle of the CB state always increases with $n$ and, even when the W state is the stable one, a metastable CB with high contact angle is observed. In other words,  when the droplet is placed on the surface initially in the CB$^0$ state, it stays in the CB phase. As a conclusion, our work shows that the contact angle in the hydrophobic state is enhanced when $n$  increases and, depending on how the droplet is placed on the surface, one can have a CB state with high contact angles even on the hydrophilic.
	
	The results shown in this section are restricted to $R_0=20\mu$m due to limitations in computational resources but it is well know that the size of the droplet affects the transition line between the W and CB states \cite{fernandes2015} and so, it is expected that the metastable states will also be affected by the increase of $R_0$. We hope to address this issue in further works.

\subsection{The influence of the fractal dimension}

On this section we discuss our simulations results in terms of the fractal dimensions, $D_f$. Several works try to establish a direct relation between the fractal dimension and wetting properties of the materials. Some argue that $D_f$ is a sufficient predictor of the contact angle \cite{gao2016tunable, piferi2021hydrophilicity} while others defend that the contact angle is independent of $D_f$ or that $D_f$ is only one of the factors impacting the contact angle \cite{yang2006influence, jain2017fractal}.

\begin{figure}[H]
	\centering
	\includegraphics[width=\linewidth]{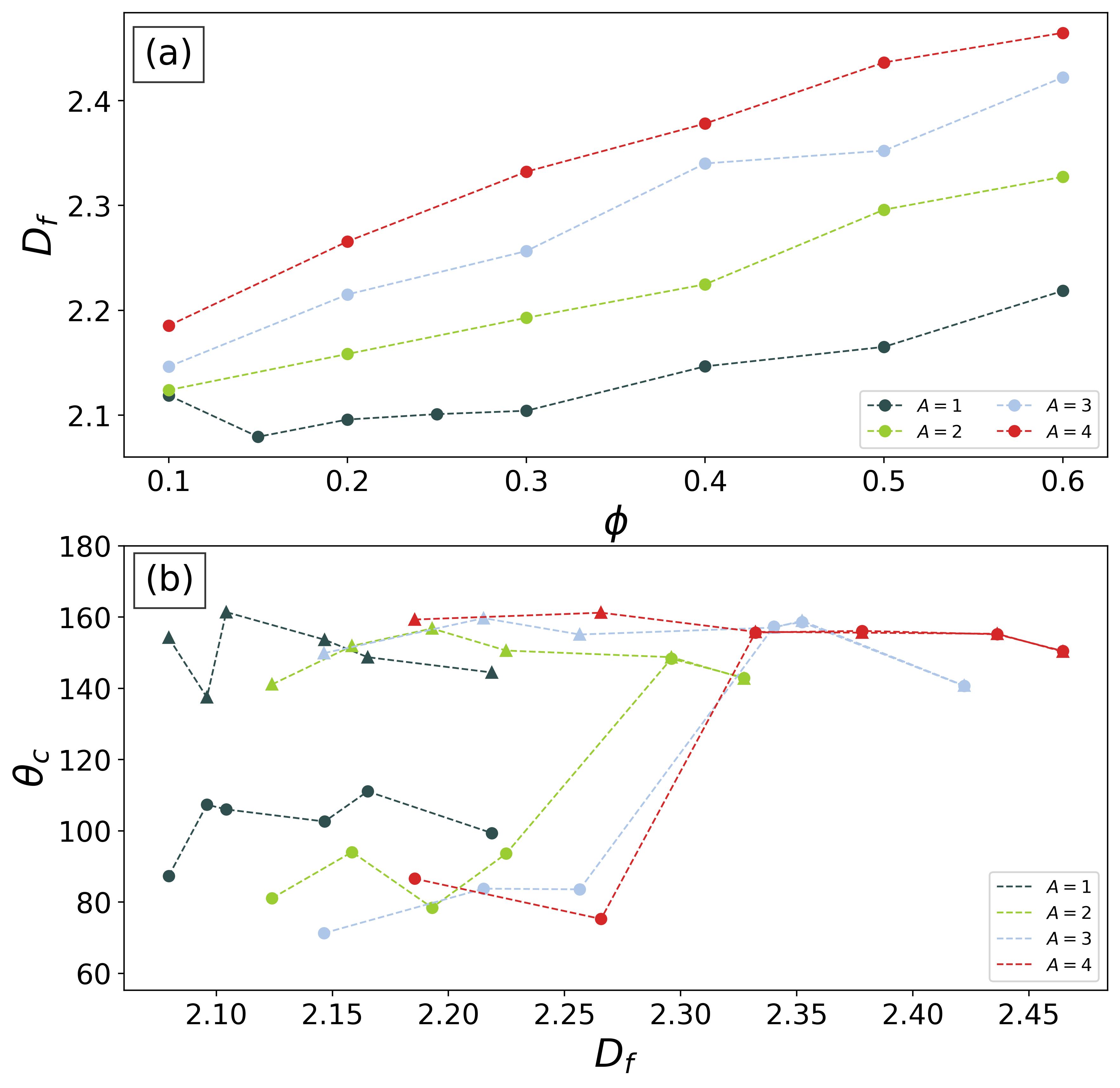}
	\caption{(a) Fractal dimensions $D_f$ as a function of $\phi$ and $A$ for $m=3$. (b) $\theta$ as a function of $D_f$ for all the simulated points. Triangles represent simulations starting in the state CB$^0$ and circles simulations starting in W$^0$. Dashed lines are guides to the eyes. }
	\label{fractal}
\end{figure}

Here we obtained the fractal dimension, $D_f = D_{cross}+1$, using a box counting method to define $D_{cross}$, the whole procedure is explained in details in the SM, similar techniques were used previously in other works \cite{Onda1996, shibuichi1998super}. Figure \ref{fractal}(a) shows $D_f$ as a function of $\phi$ and $A$ for $m=3$.

Figure \ref{fractal}(b) shows the contact angle as a function of $D_f$ for all the simulated points for auto-similarity level $n=3$. Note that the contact angles obtained when the simulation starts on state CB$^0$ are independent of $D_f$. On the other hand, when we start in state W$^0$ the contact angle starts on a low contact angle for small $D_f$ and, at a certain fractal dimension, grows to the same value as the one obtained by the simulations starting in CB$^0$. In other words, the dependence on the initial conditions decreases when $D_f$ increases.

The result suggests that the $D_f$ is not a good predictor of the contact angle of the droplet, as implied by models where the fractal dimension is used as a central parameter \cite{hazlett1990fractal, Onda1996}.

\section{Conclusions}
\label{section_conclu}

In this work we used a theoretical model and Monte Carlo simulations of the 3 spin Potts model to study the wetting properties of a particular case of a fractal surface, namely a hierarchical structured surface.

Using the theoretical model, which takes into account the global energy to create interfaces, and a minimization procedure,  we were able to predict the most favorable wetting state and the contact angle of the droplet for a range of geometric parameters. We find that for lower values of $A$ and $\phi$ the droplet is on the W state, while increasing this parameters leads to the CB state.  It is shown that the CB phase displays larger contact angle as $n$ increases, which means an enhancement of the hydrophobicity of the surface when the  auto-similarity level $n$ increases. Also, for infinite droplet radius the energy associated to the CB and W states increase with $n$  when considering a surface with Young contact angle $\theta_Y>90^\circ$. 

Theoretical results were compared with the simulations and it was observed a good agreement in regards to the energy but the the contact angle dependence on $n$ is less pronounced in simulations.  We also observed that the final state of the droplet has a strong dependency on its initial state which is associated with a metastability of the contact angle. 

This metastability was systematically evaluated for all the simulated points and observed that metastable states are more common for lower values of $A$ and $\phi$, specially in the region where the W is predicted to be the most stable state. Thus, even when in hydrophilic regions of the phase diagram one can find a droplet with high contact angle depending on its initial condition.

Finally we discussed our results in terms of the fractal dimension $D_f$ of the surfaces. It is observed a dependence of the contact angle of the droplet on its initial condition, suggesting that the fractal dimension alone cannot predict the contact angle of the droplet, as dictated by some theories \cite{Onda1996}. 

This works only takes into account a particular case of fractal  surfaces, which is very ordered. However, several biological and artificial fractal surface have some degree of disorder\cite{neinhuis1997characterization,synytska2009wetting, bhushan2009micro}. It would be nice to address disorder in a simple model like the one proposed here where it is possible to compute energy and contact angle for an infinite range of levels $n$. Also, we could easily introduce chemical and/or structure disorder and optimize these properties in order to obtain smart surfaces for different technological applications\cite{zhang2020wrinkled, wang2018pneumatic, liu2013bio}.  Another limitation here is the computational resources, which only allowed us to investigate very small droplets. It would be relevant to evaluate finite size effects in the metastability observed in this work.

\section{Supplementary material}

See supplementary material for a detailed description and an example of the energy minimization process for the continuous model, the methodology used in choosing the numerical parameters, and supplementary results.

\section*{Data Availability Statement}

The data that supports the findings of this study are available within the article and its supplementary material.

\begin{acknowledgments}
	We thank the Brazilian agency CAPES and CNPq for the financial support.  We also
	acknowledge the use of the Computational Center of the New
	York University (NYU). 
\end{acknowledgments}

%\appendix

\newpage

%\nocite{*}
\bibliography{refs}% Produces the bibliography via BibTeX.

\end{document}